\documentclass{article}

\def\bSig\mathbf{\Sigma}

\newcommand{\RNum}[1]{\uppercase\expandafter{\romannumeral #1\relax}}

\usepackage{arxiv}

\usepackage[utf8]{inputenc} 
\usepackage[T1]{fontenc}    
\usepackage{hyperref}       
\usepackage{url}            
\usepackage{booktabs}       
\usepackage{amsfonts}       
\usepackage{nicefrac}       
\usepackage{microtype}      
\usepackage{lipsum}		
\usepackage{graphicx}
\usepackage{natbib}
\usepackage{doi}

\usepackage{amsmath}
\usepackage{rotating}
\usepackage{array}
\usepackage{subcaption}
\usepackage{float}
\usepackage{placeins} 
\usepackage{multirow}
\usepackage{bm}

\usepackage{listings}
\usepackage{xcolor}
\usepackage{setspace}

\definecolor{codegray}{rgb}{0.5,0.5,0.5}
\definecolor{codepurple}{rgb}{0.58,0,0.82}
\definecolor{backcolour}{rgb}{0.95,0.95,0.95}

\lstdefinestyle{Rstyle}{
  backgroundcolor=\color{backcolour},
  commentstyle=\color{codegray},
  keywordstyle=\color{blue},
  numberstyle=\tiny\color{codegray},
  stringstyle=\color{codepurple},
  basicstyle=\ttfamily\footnotesize,
  breaklines=true,
  captionpos=b,
  keepspaces=true,
  numbers=left,
  numbersep=5pt,
  showstringspaces=false,
  tabsize=2
}
\title{Evaluating informative cluster size in cluster randomized trials}

\author{ {Bryan S. Blette} \\
	Department of Biostatistics\\
	Vanderbilt University Medical Center\\
	Nashville, TN, USA \\
    \And
    {Zhe Chen} \\
	Division of Biostatistics\\
	University of Pennsylvania \\
	Philadelphia, PA, USA \\
    \And
    {Brennan C. Kahan} \\
	Institute of Clinical Trials and Methodology\\
	MRC Clinical Trial Unit at UCL\\
	London, UK \\
    \And
    {Andrew Forbes} \\
	Division of Quantitative Research Methodology\\
	Monash University\\
	Melbourne, AU \\
    \And
    {Michael O. Harhay} \\
	Division of Biostatistics\\
	University of Pennsylvania\\
	Philadelphia, PA, USA \\
	\And
	{Fan Li}\thanks{Corresponding author: \texttt{fan.f.li@yale.edu}} \\
	Department of Biostatistics\\
	Yale University\\
	New Haven, CT, USA \\
}

\date{}

\renewcommand{\shorttitle}{\textit{arXiv} Template}

\hypersetup{
pdftitle={Evaluating informative cluster size in cluster randomized trials},
pdfsubject={q-bio.NC, q-bio.QM},
pdfauthor={Bryan S. Blette, Fan Li},
pdfkeywords={Cluster randomized trials, Cluster-average treatment effect, Individual-average treatment effect, Estimands, Informative cluster size, Simulation studies},
}

\begin{document}
\maketitle

\begin{abstract}
	In cluster randomized trials, the average treatment effect among individuals (i-ATE) can be different from the cluster average treatment effect (c-ATE) when informative cluster size is present, i.e., when treatment effects or participant outcomes depend on cluster size. In such scenarios, mixed-effects models and generalized estimating equations (GEEs) with exchangeable correlation structure are biased for both the i-ATE and c-ATE estimands, whereas GEEs with an independence correlation structure or analyses of cluster-level summaries are recommended in practice. However, when cluster size is non-informative, mixed-effects models and GEEs with exchangeable correlation structure can provide unbiased estimation and notable efficiency gains over other methods. Thus, hypothesis tests for informative cluster size would be useful to assess this key phenomenon under cluster randomization. In this work, we develop model-based, model-assisted, and randomization-based tests for informative cluster size in cluster randomized trials. We construct simulation studies to examine the operating characteristics of these tests, show they have appropriate Type \RNum{1} error control and meaningful power, and contrast them to existing model-based tests used in the observational study setting. The proposed tests are then applied to data from a recent cluster randomized trial, and practical recommendations for using these tests are discussed.
\end{abstract}


\section{Introduction}
\label{s:intro}

Cluster randomized trials (CRTs) are randomized experiments where clusters or groups of individuals are randomized to different treatments rather than randomization occurring at the individual level~\citep{murray1998design}. CRTs are a popular design choice when an intervention is naturally administered at a cluster level or when cluster randomization is more practical than individual randomization. For example, this is the case when randomizing different villages to receive vaccine or control in a vaccine trial. Although accounting for clustering remains a central consideration in planning and analyzing CRTs, there has been increasing attention to the precise definition and specification of treatment effect estimands---the target quantity that reflects the scientific question---in the context of CRTs \citep{kilpatrick2020estimands,kahan2023estimands,kahan2024demystifying,wang2024model}.

There are different estimands that can be targeted in a CRT, including those that are marginal or conditional in nature~\citep{kahan2024demystifying}, as well as those representing individual-average or cluster-average quantities~\citep{kahan2023estimands}. We focus on two popular marginal estimands on the difference scale: the individual-average treatment effect (i-ATE) and the cluster-average treatment effect (c-ATE). The former is an average of the potential outcome contrasts across all individual participants, while the latter is an average of the cluster-average effects. These two estimands are conceptually different as they assign equal weight to each individual unit or cluster unit, and thus refer to conceptually different units of inference \citep{hemming2023commentary}. In particular, the i-ATE estimand is the usual treatment effect estimand specified in an individually-randomized trial, which is often still of interest in CRTs. On the other hand, the c-ATE estimand is unique to CRTs and could be relevant when e.g., studying the impact of a cluster-level education intervention on provider prescription rates of a drug measured at the individual level. 

Mathematically, the magnitudes of i-ATE and c-ATE differ when informative cluster size (ICS) is present~\citep{kahan2023estimands}, i.e., when the potential outcomes themselves or the treatment effect contrasts depend on cluster size, subsequently referred to as Type A or Type B ICS, respectively. This phenomenon is plausible in many CRTs; for example in a hospital-based CRT, larger hospitals may have more staff and resources, be more experienced with running trials or administering the intervention, or serve fundamentally different populations than smaller hospitals; each of these aspects may associate with individual potential outcomes or the treatment effect within that cluster. For collapsible treatment effect measures, the i-ATE and c-ATE will coincide unless there is Type B ICS; for non-collapsible effect measures, the i-ATE and c-ATE will only coincide if neither type of ICS is present. While the presence of ICS should not impact the primary estimand of interest in a trial, it determines the difference between these two estimands and thus 
has important implications for the validity of certain estimators. For example, \citet{wang2022two} and \citet{kahan2024demystifying} have both demonstrated that when ICS is present, the treatment coefficient estimator from linear mixed models and generalized estimating equations with an exchangeable working correlation structure is biased for both the i-ATE and c-ATE estimands. Notably, these analysis methods constitute the majority of analyses reported in previous systematic reviews of published CRTs \citep{fiero2016statistical,offorha2022statistical}. In contrast, depending on the choice of weights, GEE with an independence working correlation structure or analyses of weighted cluster summaries are unbiased and robust to ICS. In the absence of ICS, however, 
linear mixed models lead to a consistent estimator for both i-ATE and c-ATE estimands (in this case, i-ATE and c-ATE are numerically equivalent), even under arbitrary model misspecification \citep{wang2021mixed}, and may improve efficiency over independence GEE. 
Therefore, the presence or absence of ICS can have important implications for the validity of certain estimators and remains a critical aspect that cannot be overlooked in CRTs.

Despite its importance, statistical tools for evaluating the presence of ICS in CRTs are scarce. Although in practice, 
the presence or absence of ICS could be assumed a priori based on content knowledge, 
the availability of data-driven evaluation of this key assumption will often help aid in the determination of ICS either based on completed CRTs or historical data. Such evaluations would allow researchers to characterize the likelihood of ICS across different contexts, providing key information to leverage in the design phase of new CRTs.
While our prior work provides an example graphical assessment that can be used to informally evaluate the presence of ICS \citep{kahan2024demystifying}, that approach has not leveraged auxiliary information from baseline covariates and there has been no prior evaluation of its performance. In this paper, we expand this toolbox by proposing formal hypothesis testing procedures under a null hypothesis that is a testable implication under the assumption of no informative cluster size. Specifically, we will investigate both asymptotic tests and randomization tests to assess whether i-ATE differs from c-ATE, with and without baseline covariate adjustment. This, in conjunction with the aforementioned graphical tools, will provide a suite of options to CRT analysts that can be used to assess this critical assumption. We study the performance of these tests in both simulations and an application to the Randomized Evaluation of Sedation Titration for Respiratory Failure (RESTORE) CRT to gain further insight. To facilitate practice, we implement each test in the \texttt{icstest} R package. 


\section{Methods}
\label{s:model}

\subsection{Notation and assumptions}

We focus on the setting of a parallel two-arm CRT which has $M$ clusters of potentially varying sizes $N_{i}$ for a total sample size of $N_{+}= \sum_{i=1}^{M} N_{i}$. Let $A_{i}$ equal 1 if cluster $i$ is assigned to intervention and 0 if cluster $i$ is assigned to control. We pursue the potential outcomes framework, and let $Y_{ij}(1)$ be the potential outcome for participant $j$ in cluster $i$ if the cluster is assigned to the intervention condition; similarly, we let $Y_{ij}(0)$ be the potential outcome if the cluster is assigned to the control condition. We first denote $\bm{X}_{ij}$ as the vector of all individual-level and cluster-level covariates that may be adjusted for in regression models to improve precision (excluding cluster size $N_i$), and $\bm{X}_{ij}$ is allowed to be empty. We assume the cluster-level Stable Unit Treatment Value Assumption such that $Y_{ij} = A_{i}Y_{ij}(1) + (1 - A_{i})Y_{ij}(0)$. We also assume cluster randomization ($\left\{\left(Y_{ij}(1), Y_{ij}(0),\bm{X}_{ij}; j=1,\ldots,N_i\right),N_i\right\} \perp \!\!\! \perp A_{i}$) and positivity ($0 < P(A = 1) < 1$) that are expected to hold by the study design.

\subsection{Existing model-based hypothesis tests for ICS: A review and objective}

While a handful of statistical tests exist for the presence of ICS, they were primarily derived for associational analysis. Furthermore, they tend to focus on the informativeness of cluster size in the context of an outcome model, such that they may serve as a valid test for Type A ICS but not for Type B ICS. For example, \cite{williamson2003marginal} and \cite{benhin2005mean} proposed an inverse cluster size weighted estimating equation for inference under ICS and \cite{benhin2005mean} derived a Wald test for a null hypothesis of no ICS by comparing this estimator to the standard GEE estimator. \cite{seaman2014review} noted that one could alternatively fit a mixed-effects model and test whether a coefficient for cluster size is equal to 0;  a similar procedure was used to assess ICS in a GEE model \citep{pavlou2013examination}. Finally, \cite{nevalainen2017tests} proposed an ICS test against the null of identical marginal distribution free of cluster size using a balanced bootstrap scheme.

To our knowledge, these tests have not been leveraged to test for ICS in the CRT setting. Even if they can be applied to assess ICS in CRTs, their sole focus on Type A ICS could have limited their applicability to explicitly address effect modification by cluster size under the potential outcomes framework. \cite{kahan2023informative} implemented a simulation-based procedure to evaluate ICS in the context of a specific CRT, but it was informal and did not fully leverage all baseline information to improve power. Given this context, the objective of this paper is to propose formal hypothesis testing procedures to assess ICS with a focus on Type B ICS, through leveraging randomization to attain robustness against model misspecification, and through leveraging baseline covariates $\bm{X}_{ij}$ for potential efficiency gain.

\subsection{A testable implication under ICS in CRTs}

We pursue a finite-population framework for causal inference, and define the estimands on the difference scale as
\begin{align*}
    \text{i-ATE} = \frac{1}{N_+} \sum_{i=1}^{M} \sum_{j=1}^{N_{i}} \left\{ Y_{ij}(1) - Y_{ij}(0) \right\},~~~~
    \text{c-ATE} = \frac{1}{M} \sum_{i=1}^{M} \left\{ \overline{Y}_{i\bullet}(1) - \overline{Y}_{i\bullet}(0) \right\}
\end{align*}
where $\overline{Y}_{i\bullet}(a) = \frac{1}{N_{i}} \sum_{j=1}^{N_{i}} Y_{ij}(a)$ for $a = 0, 1$ is the cluster-specific mean potential outcome. By definition, the i-ATE is an average of potential outcome contrasts over the entire individual population, the same estimand frequently used in individually randomized trials without any clustering structure. On the other hand, the c-ATE is the average treatment effect across clusters. As a key implication of ICS is that the magnitude of these two estimands differ \citep{kahan2021estimands,kahan2024demystifying,kahan2023informative}, our goal is to develop a test under a null hypothesis of $H_{0}: \text{i-ATE} = \text{c-ATE}$. To accomplish this, we first note that
\begin{align*}
    \Delta = \text{i-ATE} - \text{c-ATE} 
    = \sum_{i=1}^{M} \left( \frac{N_{i}}{N_+} - \frac{1}{M} \right) \left\{ \overline{Y}_{i\bullet}(1) - \overline{Y}_{i\bullet}(0) \right\}
    = \sum_{i=1}^{M} \pi_{i} \left\{ \overline{Y}_{i\bullet}(1) - \overline{Y}_{i\bullet}(0) \right\}
\end{align*}
which has the form of a weighted average treatment effect estimand with cluster-specific weights equal to $\pi_{i}$. This motivates us to adapt existing estimation procedures for the weighted average treatment effect estimand $\Delta$ into valid and efficient hypothesis tests for $H_{0}: \sum_{i=1}^{M} \pi_{i} \left\{ \overline{Y}_{i\bullet}(1) - \overline{Y}_{i\bullet}(0) \right\} = 0$, as a data-driven approach for assessing ICS.


Several remarks are to follow regarding the weight $\pi_{i}$ in $\Delta$. First, in the trivial case where a CRT does not have varying cluster sizes and ICS is irrelevant, then $N_{i} = N_+/M$ for all $i$ such that $\pi_{i} = 0$ for all $i$ and $\Delta=\text{i-ATE}-\text{c-ATE}=0$. Second, the sum of the weights $\sum_{i=1}^{M} \pi_{i} = 0$. Third, these weights can be positive or negative, and in fact outside of the trivial case, there will always be at least one negative weight. This has implications for the practical implementation of proposed hypothesis tests. The weight for a specific cluster can be interpreted heuristically as proportional to the deviation of the cluster's size from the average cluster size. Clusters which are much larger (smaller) than average will have a large positive (negative) weight and clusters close to the average size will have a weight closer to 0. Finally, upon closer inspection, the null hypothesis $H_0$ can be further expressed as
\begin{align*}
0=\frac{N_+}{M}\Delta&=\frac{1}{M}\sum_{i=1}^{M} N_i \left\{ \overline{Y}_{i\bullet}(1) - \overline{Y}_{i\bullet}(0) \right\}-\frac{N_+}{M} \times\frac{1}{M}\sum_{i=1}^{M} \left\{ \overline{Y}_{i\bullet}(1) - \overline{Y}_{i\bullet}(0) \right\}\\
&=\mathcal{COV}\left\{N_i, \left\{ \overline{Y}_{i\bullet}(1) - \overline{Y}_{i\bullet}(0) \right\}\right\},
\end{align*}
where $\mathcal{COV}$ is the finite-population covariance operator. This re-expression delivers an important insight that testing $H_0:\Delta=0$ is equivalent to testing zero covariance between cluster size the unobserved cluster-specific treatment effect; in other words, the test for ICS is to formally assess whether the cluster size $N_i$ is a cluster-level effect modifier. 

Here and throughout, we make a minor abuse of terminology in that for collapsible treatment effect measures such as the difference scale we consider, this is a testing framework of just Type B ICS, i.e., the case when specifically the treatment effect depends on cluster size, since i-ATE and c-ATE will coincide if only the outcomes but not the treatment effect depend on cluster size. However, we broadly refer to the proposed tests below as tests of informative cluster size for ease of communication and in line with prior literature. For non-collapsible effect measures (such as the odds ratio), the null hypothesis of $H_{0}: \text{i-ATE}=\text{c-ATE}$ has a one-to-one correspondence with the null hypothesis of no informative cluster size (regardless of ICS type). Since whether or not i-ATE and c-ATE coincide is the key factor in estimator validity in this setting, we propose that this testing framework is more relevant to our research question than the tests of only Type A ICS previously described. Finally, we center our primary discussion around estimands on the difference scale, and we discuss possible future work to address ratio estimands in Section \ref{s:discuss}.

\subsection{Model-assisted Hypothesis Tests for ICS}

\cite{su2021model} derived a series of estimators for weighted average treatment effects in cluster randomized trials under the design-based perspective and evaluated their performance in a variety of scenarios. They considered both weighted ordinary least squares (OLS) regression of cluster means as well as OLS regression of weighted cluster means. Because (i) they showed that the latter approach has lower asymptotic variance than the former approach and (ii) the former approach will have implementation concerns in the presence of negative weights for many software packages, we propose two hypothesis testing approaches (unadjusted and covariate-adjusted) using OLS of weighted cluster means catered to our ICS null hypothesis. While explicit or implicit negative weights have introduced concerns in estimating weighted average treatment effects in certain settings~\citep{small2017instrumental}, recent work has shown that OLS regression under a design-based framework avoids these issues~\citep{borusyak2024negative}.

\textbf{Unadjusted model-assisted ICS test:} Let $\overline{Y}_{i\bullet} = \frac{1}{N_{i}} \sum_{j=1}^{N_{i}} Y_{ij}$ and let $\widetilde{Y}_{i\bullet} = M\pi_{i} \overline{Y}_{i\bullet}$. Consider the regression model $\widetilde{Y}_{i\bullet} = \beta_{0} + \beta_{1}A_{i} + \epsilon_{i}$ such that an unbiased estimator of $\Delta$ is given by the usual OLS estimator
\begin{equation*}
    \widehat{\Delta} = \widehat{\beta}_{1} = \frac{\sum_{i=1}^{M} (A_{i} - \overline{A})\widetilde{Y}_{i\bullet}}{\sum_{i=1}^{M} (A_{i} - \overline{A})^{2}}
\end{equation*}
Then a standard $t$-test for null $H_{0}: \beta_{1} = 0$ serves to assess the existence of ICS, i.e., using the Wald test statistic, $\widehat{T} = \widehat{\Delta} / \widehat{se}(\widehat{\Delta}) \sim t(M-1)$, 
where $t(M-1)$ notates a $t$-distribution with $M-1$ degrees of freedom. Importantly, when the number of clusters is small, a standard $t$-distribution may be inappropriate and a $t$-distribution with corrected degrees of freedom may be more suitable~\citep{kenward1997small}. The Huber-White standard error is necessary for consistent estimation.

\textbf{Covariate-adjusted model-assisted ICS test:} Covariate adjustment can improve statistical power in many contexts. Let $\vec{\bm C_{i}}$ be a row vector of cluster-level covariates that are potentially informative of the outcome for cluster $i$ and let $\overline{\vec{\bm C}} = \frac{1}{M}\sum_{i=1}^{M} \vec{\bm C_{i}}$ be a corresponding row vector of cluster averages. Let $\bm{1}, \bm{A}, \bm{C}$, and $\overline{\bm{C}}$ be a set of column vectors corresponding to the intercept, treatment, and covariates for each cluster and let $\bm{X} = (\bm{1}, \bm{A}, \bm{C} - \overline{\bm{C}}, \bm{A}\circ(\bm{C} - \overline{\bm{C}}))$ be an $M \times (2+2p)$ design matrix, where $\circ$ denotes the element-wise product applied row-wise. Consider the regression model $\bm{\widetilde{Y}} = \bm{\beta} \bm{X} + \bm{\epsilon}$ where $\bm{\widetilde{Y}}$ is a column vector with $ith$ element $\widetilde{Y}_{i.}$ such that an unbiased estimator of $\bm{\beta}$ is given by the OLS estimator, $\bm{\widehat{\beta}} = (\bm{X}^{T}\bm{X})^{-1}\bm{X}^{T}\bm{\widetilde{Y}}$. 
Then $\widehat{\Delta}^{adj}$ is given by the second element of $\bm{\widehat{\beta}}$. The mean-centering of covariates is not necessary to improve power, but allows for immediate interpretation of this element of $\bm{\widehat{\beta}}$ as a marginal effect (when using non-centered covariates, an additional marginalization step must be pursued). A similar testing procedure to the unadjusted approach can be leveraged with $\widehat{T}^{adj} = \widehat{\Delta}^{adj} / \widehat{se}(\widehat{\Delta}^{adj}) \sim t(M - 2p - 1)$. 

Covariate adjustment has been shown to improve power in many settings, although improvement is not guaranteed, especially when the sample size (here the number of clusters) is small. Importantly, when performing linear regression with standardization (which this proposed test leverages), covariate adjustment does not asymptotically bias the estimate of the treatment effect when the treatment is randomized even when the model specification is incorrect~\citep{van2024covariate}. Hence, this ``model-assisted'' approach may be fairly robust in practice, particularly for CRTs with a large number of clusters. However, caution should be exercised when adjusting for covariates in a CRT with a small number of clusters, as the gain in power and precision may not outweigh the reduction in degrees of freedom. There are other estimators of weighted average treatment effects that can potentially be leveraged in a model-assisted analysis to test for ICS (e.g., using cluster total summary statistics rather than cluster averages), but these exhibit similar performance in other settings~\citep{su2021model}, so we exclude them for brevity. Adjustment for mean-centered cluster size as an additional covariate is recommended in practice.

\subsection{Randomization-based Hypothesis Tests for ICS}

\cite{wu2021randomization} showed that, under certain conditions, the Fisher Randomization Test (FRT) constructed under a sharp null hypothesis remains asymptotically valid (albeit potentially conservative) for testing a compatible weak null hypothesis in randomized experiments. Crucially, this validity hinges on employing a test statistic whose randomization distribution stochastically dominates its sampling distribution asymptotically under the weak null. For our weak null hypothesis of interest, $H_{0}: \text{i-ATE} = \text{c-ATE}$, the test statistic from the model-assisted approach using the Huber-White robust standard error falls into the class of statistics with this property, as discussed in Section 3.3 of \citet{wu2021randomization}.

\textbf{Unadjusted randomization-based ICS Test:} Leveraging the above insights, we propose a randomization-based test for ICS as follows. 
To facilitate testing the weak null hypothesis $H_{0}: \text{i-ATE} = \text{c-ATE}$, we consider the compatible sharp null hypothesis
\begin{equation}\label{sharp null}
    H_{0F}: \pi_{i} \overline{Y}_{i.}(0) = \pi_{i} \overline{Y}_{i.}(1), \quad \text{for}~ i = 1, \dots, M,
\end{equation}
which asserts that each cluster’s weighted average potential outcomes under treatment and control are equal. Under $H_{0F}$, all cluster mean potential outcomes equal their observed outcomes $\widetilde{Y}_{i.}$, and thus the FRT reduces to a classical permutation test at the cluster level.
Specifically, we first compute the test statistic for the observed data $\widehat{T}_{obs}$. Then permute the treatment assignments at the cluster level $D$ times and calculate the test statistic under each permutation,  $\widehat{T}_{d}$, for $d=1,...,D$. The randomization-based $p$-value for testing $H_{0}$ is then given by the proportion of permuted test statistics which are at least as extreme as the observed test statistic: $\frac{1}{D} \sum_{d=1}^{D} I(|\widehat{T}_{d}| \geq |\widehat{T}_{obs}|)$. While all possible permutations can be enumerated when the number of clusters is very small, this quickly becomes computationally impractical as $M$ increases. In practice, a random subset of $D$ permutations is typically used, with $D = 5000$ recommended for tests at level $\alpha = 0.05$~\citep{marozzi2004some}. For smaller $\alpha$ levels, larger values of $D$ are required to maintain adequate precision of the $p$-value.

Notably, the above randomization test not only provides asymptotically valid inference for the weak null $H_{0}$, but also remains finite-sample exact for testing the sharp null in \eqref{sharp null}. Moreover, although we employ regression estimators based on OLS as test statistics, the validity of the resulting $p$-values does not depend on correct model specification and is justified by cluster randomization.

\textbf{Covariate-adjusted randomization-based ICS Test:} While \citet{wu2021randomization} established asymptotic validity of randomization-based tests for weak null hypotheses, their results do not explicitly address covariate adjustment. Follow-up work described in \cite{zhao2021covariate} showed that studentized statistics from covariate-adjusted models are similarly valid under certain conditions for testing the weak null hypothesis of zero average treatment effect, but more general weighted average treatment effects were not considered. Nonetheless, motivated by the potential gains in efficiency and power from covariate adjustment, we explore this approach empirically for our null hypothesis. 

The proposed covariate-adjusted procedure for ICS follows the same randomization framework described above, but employs the test statistic from a covariate-adjusted model. This follows the recommended ``model-output'' strategy defined by \cite{zhao2021covariate}. Specifically, we compute the $p$-value as
$p^{\text{adj}} =\frac{1}{D} \sum_{d=1}^{D} I(|\widehat{T}^{adj}_{d}| \geq |\widehat{T}^{adj}_{obs}|),$
where $\widehat{T}^{adj}_{obs}$ and $\widehat{T}^{adj}_{d}$ denote the covariate-adjusted model-assisted test statistics for the observed data and the $d$-th permutation, respectively. While formal theoretical analysis of the asymptotic properties for the covariate-adjusted randomization-based test are beyond the scope of this work, we evaluate its empirical performance through simulation studies and real data analysis to inform its potential practical use in ICS settings.

\subsection{An Extension of the Simple Model-Based Approach}

We additionally describe the unadjusted and covariate-adjusted model-based ICS tests that are derived from the associational analysis literature. 
We consider the method discussed earlier from \cite{seaman2014review} and \cite{pavlou2013examination}, where one tests whether a coefficient for cluster size is equal to 0 in a GEE or mixed-effects outcome model. This should be a valid test of Type A ICS, but our collapsible estimand only requires consideration of Type B ICS. We consider a slight extension of this method where one fits an outcome model with an interaction between treatment and cluster size and tests whether the corresponding model coefficient is equal to 0. Similarly, one can specify a series of nonlinear interaction terms and perform a chunk test for whether an overall interaction is present. As with the other two approaches, one can further adjust for covariates in the model to potentially improve power. The utility of such a test will be compared with the model-assisted and randomization-based ICS tests in the subsequent simulation studies. 

\subsection{Practical Considerations}

We have described three different frameworks for testing for ICS in CRTs, each of which can be adjusted for covariates or unadjusted. Table 1 summarizes the resulting 6 proposed testing procedures along with the existing approaches from the associational analysis setting. We acknowledge that formal hypothesis testing often puts overemphasis on prespecified thresholds, such as the classic $\alpha = 0.05$ boundary. Scientific conclusion should not be based solely on a p-value but on the broader context of the problem and its investigation~\citep{wasserstein2016asa}. To that note, broader inference of the difference quantity $\text{i-ATE} - \text{c-ATE}$ can be used to augment these tests for more holistic evaluation of ICS. While not the focus of this work, confidence intervals for $\text{i-ATE} - \text{c-ATE}$ can be derived directly from the model-assisted approach and indirectly from the randomization-based test through test inversion~\citep{rabideau2021randomization}.

\renewcommand{\arraystretch}{1.5} 
\begin{table}[h]
\centering
\begin{tabular}{p{5cm} p{9cm}}
\toprule
\multicolumn{2}{c}{\textbf{Existing Methods}} \\
\midrule
\makebox[5cm]{\textbf{Method}} & \makebox[9cm]{\textbf{Short Description}} \\
\midrule
Behnin et al. (2005) & Wald test comparing test statistic under ICS to test statistic under no ICS (Type A) \\
Pavlou et al. (2013) & Test coefficient for cluster size in GEE model \\
Seaman et al. (2014) & Test coefficient for cluster size in mixed-effects model \\
Nevalainen et al. (2017) & Balanced bootstrap approach \\
\midrule
\multicolumn{2}{c}{\textbf{Proposed Methods}} \\
\midrule
\makebox[5cm]{\textbf{Method}} & \makebox[9cm]{\textbf{Short Description}} \\
\midrule
Unadjusted Model-Assisted Test & Test coefficient for treatment in the OLS of weighted cluster means, where weights are proportional to deviations of each cluster size from the average cluster size \\
Adjusted Model-Assisted Test & Test coefficient for treatment in the OLS fit of weighted cluster means conditional on the centered cluster-mean covariates and their interactions with treatment \\
Unadjusted Randomization-Based Test & Permute the cluster-level treatments and compute the unadjusted model-assisted test statistic for each permutation. The p-value is the proportion of permuted test statistics that are as or more extreme than the observed test statistic \\ 
Adjusted Randomization-Based Test & Permute the cluster-level treatments and calculate the adjusted model-assisted test statistic for each permutation. The p-value is the proportion of permuted test statistics that are as or more extreme than the observed test statistic \\
Unadjusted Model-Based Test & Test coefficient(s) for interaction(s) between treatment and cluster size in a GEE outcome model \\
Adjusted Model-Based Test & Test coefficient(s) for interaction(s) between treatment and cluster size in a GEE outcome model, adjusting for mean-centered covariates \\
\bottomrule
\end{tabular}
\caption{Description of existing methods from the observational study literature and proposed methods specific to the difference estimand and CRT setting. The phrase ``test coefficient'' refers to a test under a null hypothesis that the coefficient is equal to 0 (or that a chunk test of relevant coefficients are equal to 0). All weighted approaches use the weights derived in Section 2.3.}
\label{table:methods}
\end{table}

As discussed when motivating tests for ICS, the presence or absence of ICS has important implications for validity of a chosen estimator. If ICS is present, a mixed-effects regression model or GEE with exchangeable correlation will be biased for both the i-ATE and c-ATE. However, two-stage procedures that select an estimator conditional on results from an initial test typically incur bias, Type \RNum{1} error, or other issues~\citep{freeman1989performance,rochon2012test,kahan2013bias}. Thus, we do not endorse using the proposed ICS tests to select between a set of estimators. In the simulation study, we will explore the implication of an intermediate test of ICS on bias and Type \RNum{1} error for the primary trial estimand/hypothesis to investigate this further, and to caution the use of such a test.

To be more specific, the proposed tests may be used in a variety of other scenarios. First, they can be used as a part of a set of post-hoc assumption checks after a pre-specified analysis is complete, when statistical assumptions are often assessed to provide a concrete context that can help interpret the primary analysis results. Second, they can be used to empirically investigate the occurrence and frequency of ICS in completed CRTs, as the prevalence of this phenomenon is currently not sufficiently understood. Third, they can be used in trial planning and drafting of statistical analysis plans when relevant pilot or earlier data are available. For example, if one is writing a statistical analysis plan for a future CRT, an ICS test reported from a previous trial with similar outcome and intervention may be useful to inform whether ICS will be expected in the future trial, and thus can inform which estimator should be specified for the analysis.

Importantly, these tests should not be the sole evidence for evaluating ICS. ICS inherently depends on the study question, type of clusters, scientific context, outcome of interest, and other facets. Graphical assessments of ICS could be used in conjunction with the proposed tests, and p-values should be interpreted holistically rather than adhering to an $\alpha = 0.05$ threshold (which is likely overly stringent, especially for CRTs with fewer clusters and lower power to detect ICS). As outlined in the following section, covariate adjustment will be critical to have meaningful power to assess ICS in most CRTs. Further, the randomization-based tests may be more appropriate to assess ICS when a CRT has fewer clusters, as is common in practice.

\section{Simulation Study}
\label{s:inf}

\subsection{Simulation 1: Establishing validity of the proposed unadjusted tests}

In the first simulation study, we compare the Type \RNum{1} error and power of the proposed unadjusted ICS tests to an existing approach used in observational studies, namely a model-based approach which tests the coefficient for cluster size in a GEE outcome model~\citep{pavlou2013examination}. As discussed earlier, this comparator is likely only a valid test for Type A ICS and not for Type B ICS.

Data were generated such that Type B ICS was present with varying magnitude, including scenarios of small, large, and no ICS, such that both power and Type I error could be thoroughly explored. First, the number of clusters was set to vary in $\{ 30, 60, 100\}$. Then the cluster sizes were generated assuming exactly half came from a Uniform(20, 50) distribution, and the other half from a Uniform (50, 80) distribution, rounded to the nearest whole number. For each individual, a potential outcome under the control condition was generated as $Y(0) \sim I(CS > 50) + N(0, 1) + \alpha_{0}$, where $I(CS > 50)$ is the indicator that the individual belongs to a cluster of size greater than 50 and $\alpha_{0}$ was a cluster-level random intercept generated such that the outcome ICC was equal to 0.1. A potential outcome under the treatment condition was generated as $Y(1) \sim 0.5 + I(CS > 50)\times(k-4) + N(0, 1) + \alpha_{0}$ for scenarios of $k \in \{1,2,...,9\}$, where $k$ is an arbitrary parameter leveraged to vary the magnitude of ICS. In particular, for $k \in \{1, 2, ..., 9\}$ there is an approximate corresponding $\text{i-ATE} - \text{c-ATE} \in \{-0.6, -0.45, -0.3, -0.15, 0, 0.15, 0.3, 0.45, 0.6\}$ where the null scenario of 0 is exact.

Then, each of the proposed unadjusted tests (model-assisted, randomization-based, and model-based) were used to test for ICS at the $\alpha = 0.05$ level, along with the comparator test. To inspect sensitivity to model misspecification, the model-based test was performed two ways, once with the correct ICS structure (that the treatment effect interacts with $I(CS > 50)$) and once with an incorrect structure (that the treatment effect interacts with the log of the cluster size). For the randomization based test, a value of $D$ must be specified. While $D = 5000$ is recommended for an individual application of a randomization-based test with $\alpha = 0.05$, this can be computationally burdensome. Previous work has highlighted that $D=500$ is likely sufficient to assess Type \RNum{1} Error of a permutation test in a simulation~\citep{marozzi2004some}; thus, we use this value throughout all simulations.

For each method and data generation scenario, our metric of interest is the probability of rejecting the null hypothesis across 2,000 simulation iterations. This is equivalent to the Type \RNum{1} error when $k = 5$ and to statistical power for all other $k$ values. With 2,000 simulations, for a test with true 5\% Type \RNum{1} error, the Monte Carlo standard error would be just under 1\% such that about 95\% of estimated Type \RNum{1} error results would fall between 4\% and 6\%.

\begin{sidewaysfigure}
    \centering
    \includegraphics[width=1\linewidth]{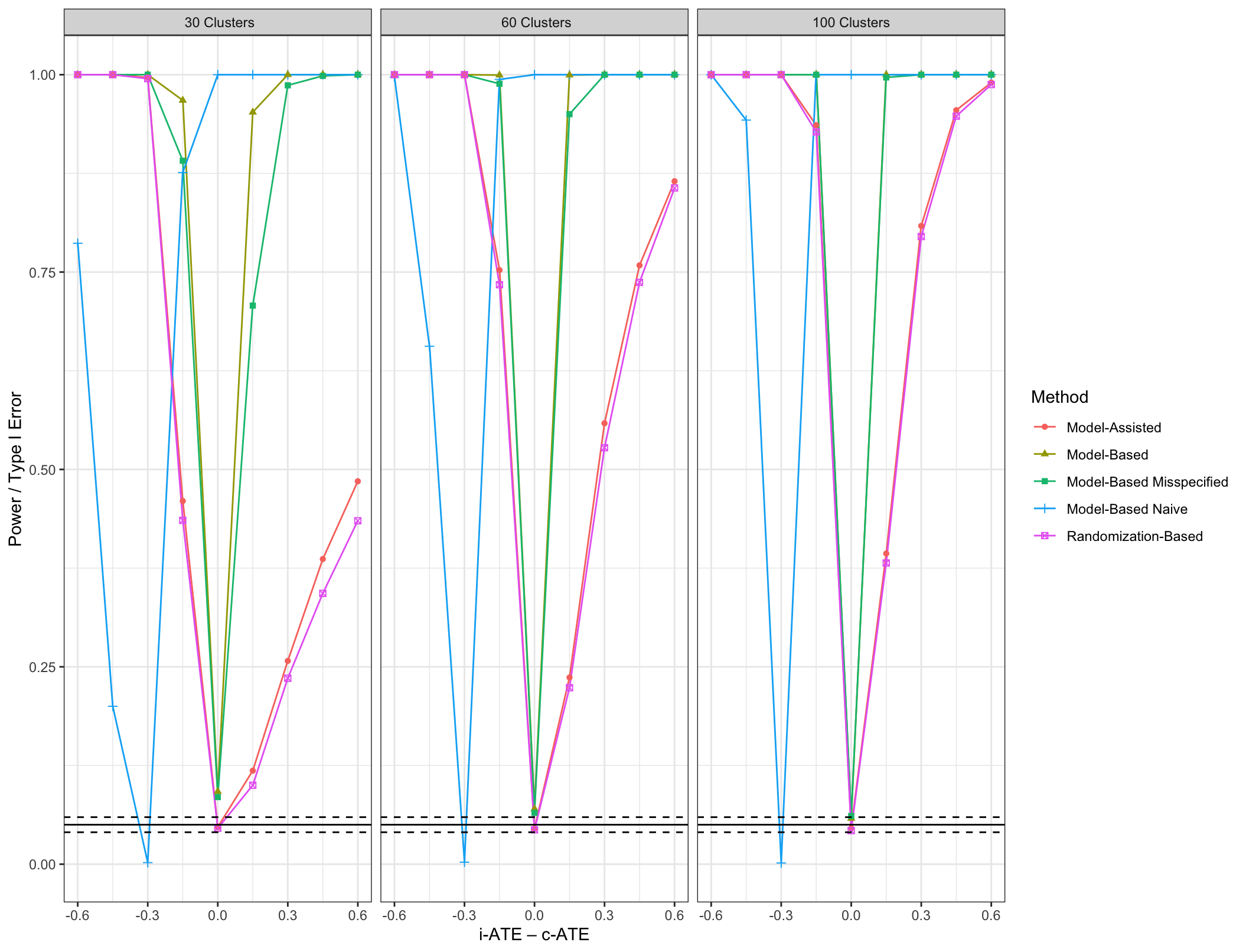}
    \caption{Results from Simulation 1. Model-Based Naive indicates the method described by \citep{pavlou2013examination}. The dashed lines indicate Monte Carlo standard error for 2000 simulations assuming a true Type I error of 5\%; this range is relevant for the middle of each panel where the null hypothesis holds.}
    \label{fig:fig1}
\end{sidewaysfigure}

The results are presented in Figure 1. The naive existing approach performs poorly in all scenarios considered. This is not surprising as it was derived to test for Type A ICS. However, this highlights the contribution of our proposed testing procedures which test for the presence of Type B ICS, which is crucial to inform estimator validity as described earlier. The model-based approach has valid Type \RNum{1} error and high power across the scenarios considered as long as the model is correctly specified. However, when the model is misspecified, it can have unacceptably high Type \RNum{1} error (about 10\% in the scenario with 30 clusters). On the other hand, the model-assisted and randomization-based tests had valid Type \RNum{1} error in all scenarios considered, but often had low power, particularly in the scenario with 30 clusters. The model-assisted and randomization-based tests had similar performance to each other, although the latter is more computationally burdensome. Results for data generated under ICC = 0.01 are presented in the Supplementary Web Material.

\subsection{Simulation 2: Investigating covariate adjustment and model robustness}

The model-assisted and randomization-based tests would be more useful in practice if they had higher power. In this simulation, we compare the performance of these tests when they are unadjusted versus when they are adjusted for an informative cluster-level covariate. We further compare to an adjusted approach which is in a sense ``misspecified''.

The simulation is constructed to explore method performance under varying Type B ICS, as before. The number of clusters, cluster sizes, and treatment assignments are generated in the same way as Simulation 1. Then a cluster-level covariate was generated as $C \sim N(0, 1)$. For each individual, a potential outcome under the control condition was generated as $Y(0) \sim I(CS > 50) + C + N(0, 1) + \alpha_{0}$, where $\alpha_{0}$ was a cluster-level random intercept generated such that the outcome ICC was equal to 0.1. A potential outcome under the treatment condition was generated as $Y(1) \sim 0.5 + C + I(CS > 50)\times (k-4) + N(0, 1) + \alpha_{0}$ for scenarios of $k \in \{1,2,...,9\}$. Thus, the magnitude of ICS will shift across values of $k$ in the same fashion as Simulation 1.

For each scenario, six hypothesis tests were performed at the $\alpha = 0.05$ level. The unadjusted model-assisted test was compared to an adjusted model-assisted test with correct specification (including $C$ as a covariate that also interacts with treatment) and an adjusted test with incorrect specification (including $C^{2}$ as a covariate that also interacts with treatment). Note that both $C$ and $C^{2}$ are already mean-centered, and that both models also correctly adjusted for cluster size. The unadjusted randomization-based test was also used and compared to two covariate adjusted versions (corresponding to using the test statistics from the correctly and incorrectly adjusted model-assisted tests).

\begin{sidewaysfigure}
    \centering
    \includegraphics[width=1\linewidth]{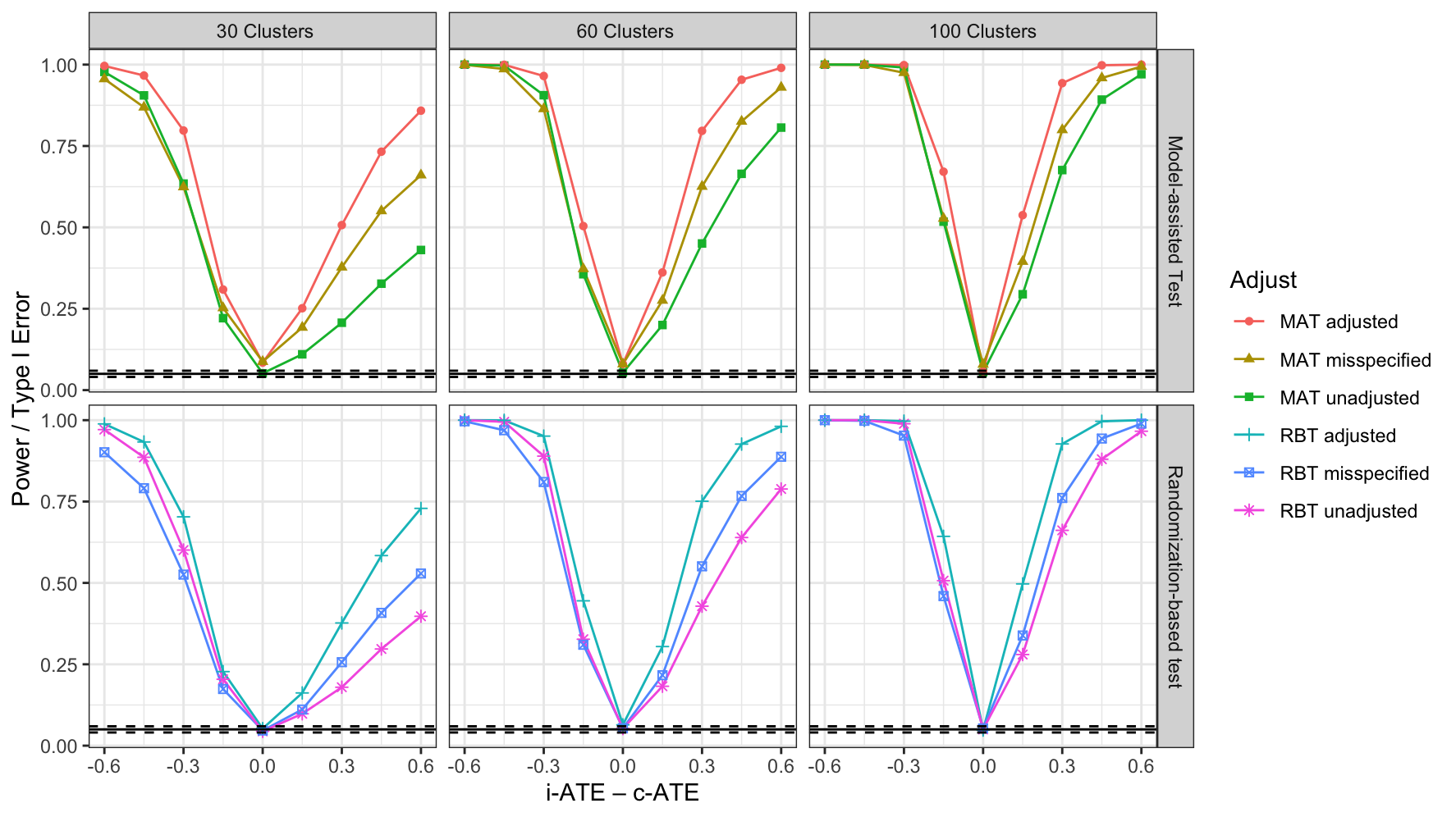}
    \caption{Results from Simulation 2. MAT = model-assisted test. RBT = randomization-based test. The dashed lines indicate Monte Carlo standard error for 2000 simulations assuming a true Type I error of 5\%; this range is relevant for the middle of each panel where the null hypothesis holds.}
    \label{fig:fig2}
\end{sidewaysfigure}

The results are presented in Figure 2. For both frameworks, the adjusted tests which leveraged correctly-specified models had uniformly higher power than the unadjusted tests. Power gains were fairly substantial, but power gains will be related to covariate informativeness in general. The approaches were slightly robust to model misspecification, but notably the adjusted misspecified tests were not uniformly more powerful than the unadjusted tests in the scenarios considered. For the scenario with 30 clusters, the model-assisted tests had Type \RNum{1} Error inflation when adjusting for covariates, however, the randomization-based tests had nominal Type \RNum{1} Error. This indicates that an adjusted randomization-based test may be a good choice for CRTs with a small number of clusters. An additional assessment of Type \RNum{1} error for the adjusted model-assisted test across varying numbers of clusters is presented in the Supplementary Web Material, along with results for data generated under ICC = 0.01.

\subsection{Simulation 3: Implications of using an intermediate ICS test in a two-stage process}

In the third simulation, we consider the implication of using an ICS test as an intermediate test to inform estimator selection in a two-stage procedure. In other settings, similar two-stage procedures can introduce notable bias and Type I error~\citep{rochon2012test}, and we do not recommend using the proposed tests for estimator selection in this manner. Rather, the goal of this simulation is to explore the extent to which a two-stage procedure would introduce bias and Type I error to motivate this recommendation in practice.

Consider two hypothetical analysts of a CRT who both want to estimate the i-ATE as their target estimand. Analyst 1 assumes there is ICS present and uses GEE with independence correlation to estimate the i-ATE. Analyst 2 first performs a model-assisted ICS test and uses GEE with independence correlation if the test is rejected at the $\alpha = 0.05$ level (and uses exchangeable correlation otherwise). As before, we consider the setting of Type B ICS, and to facilitate exploration of Type I error, data were generated such that the true i-ATE was equal to 0, but the c-ATE (and magnitude of ICS) are varied.

In particular, the number of clusters was set to 100 to explore large-sample performance of this procedure. Then the cluster sizes were generated assuming exactly half came from a Uniform(5, 20) distribution, and the other half from a Uniform (20, 35) distribution, rounded to the nearest whole number. For each individual, a potential outcome under the control condition was generated as $Y(0) \sim I(CS > 20) + C + N(0, 1) + \alpha_{0}$, where $I(CS > 20)$ is the indicator that the individual belongs to a cluster of size greater than 20, $C$ is a cluster level covariate generated as $N(0, 1)$, and $\alpha_{0}$ was a cluster-level random intercept generated such that the outcome ICC was equal to 0.1. A potential outcome under the treatment condition was generated as $Y(1) \sim I(CS > 20)\times\{1 + 0.1(k-1)\} - I(CS \leq 20)\times 0.1(k-1)\times S + C + N(0, 1) + \alpha_{0}$ for scenarios of $k \in \{1,2,...,9\}$, where $S$ is equal to the ratio of the number of clusters larger than 20 divided by the number of clusters with 20 or fewer individuals (which ensures the $\text{i-ATE}$ remains fixed at 0). Thus, there is no ICS when $k=1$, but the absolute value of $\text{c-ATE}$, and thus the magnitude of ICS, will increase as $k$ increases. For estimation, all GEE models and the intermediate ICS test adjusted for the informative covariate appropriately. The metrics of interest were bias in estimating the i-ATE across the range of ICS magnitudes, as well as the average Type I error of each full estimation procedure.

\begin{sidewaysfigure}
    \centering
    \includegraphics[width=1\linewidth]{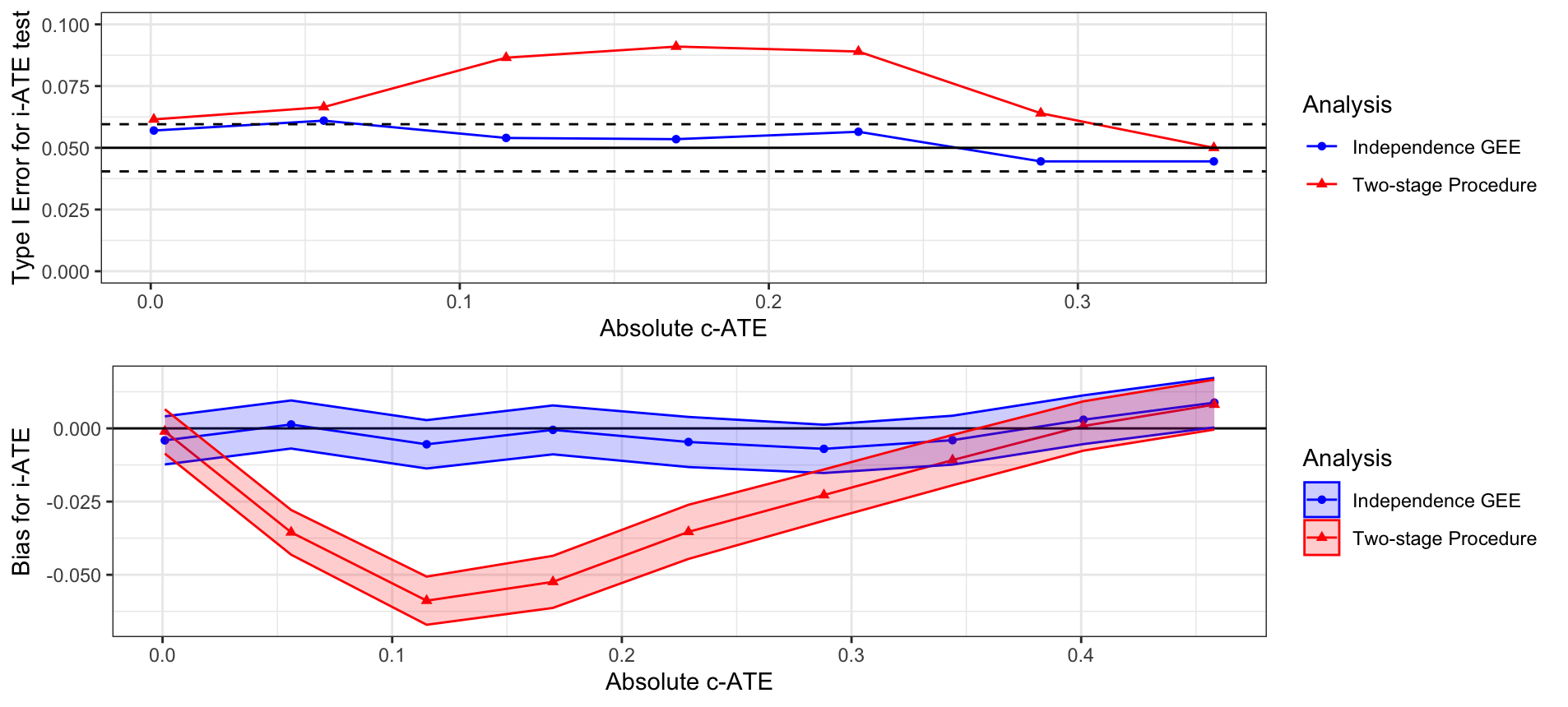}
    \caption{Results from Simulation 3 comparing an analyst who performs a two-stage procedure to an analyst who uses independence estimating equations. In the top panel, Type I error is reported and the dashed lines indicate Monte Carlo standard error for 2000 simulations assuming a true Type I error of 5\%. In the bottom panel, bias is reported with corresponding Monte Carlo confidence intervals to quantify uncertainty from using a finite number of simulations.}
    \label{fig:fig3}
\end{sidewaysfigure}

Results are presented in Figure 3. As expected, only Analyst 1 who uses independence GEE is approximately unbiased across the scenarios considered. Analyst 2 uses a two-stage procedure which is unbiased and has valid Type I error when there is no ICS present, or when the magnitude of ICS is so strong that rejection of the ICS test null hypothesis is nearly guaranteed (in which case, they essentially become Analyst 1). However, the two-stage procedure is biased and has inflated Type I error under mild to moderate ICS, with both mirroring a quadratic relationship to ICS magnitude. Thus, as in other fields, we do not recommend using a two-stage estimation procedure in practice. Results for data generated under ICC = 0.01 are presented in the Supplementary Web Material.


\section{Application to the RESTORE Cluster Randomized Trial}

We applied our proposed methods to data from the Randomized Evaluation of Sedation Titration for Respiratory Failure (RESTORE) study, a multicenter cluster randomized trial evaluating a nurse-implemented, goal-directed sedation protocol in mechanically ventilated children with acute respiratory failure due to lower respiratory tract disease \citep{curley2015protocolized}.

\begin{figure}[H]
    \centering
    \begin{subfigure}[t]{0.48\textwidth}
        \centering
        \includegraphics[width=\textwidth]{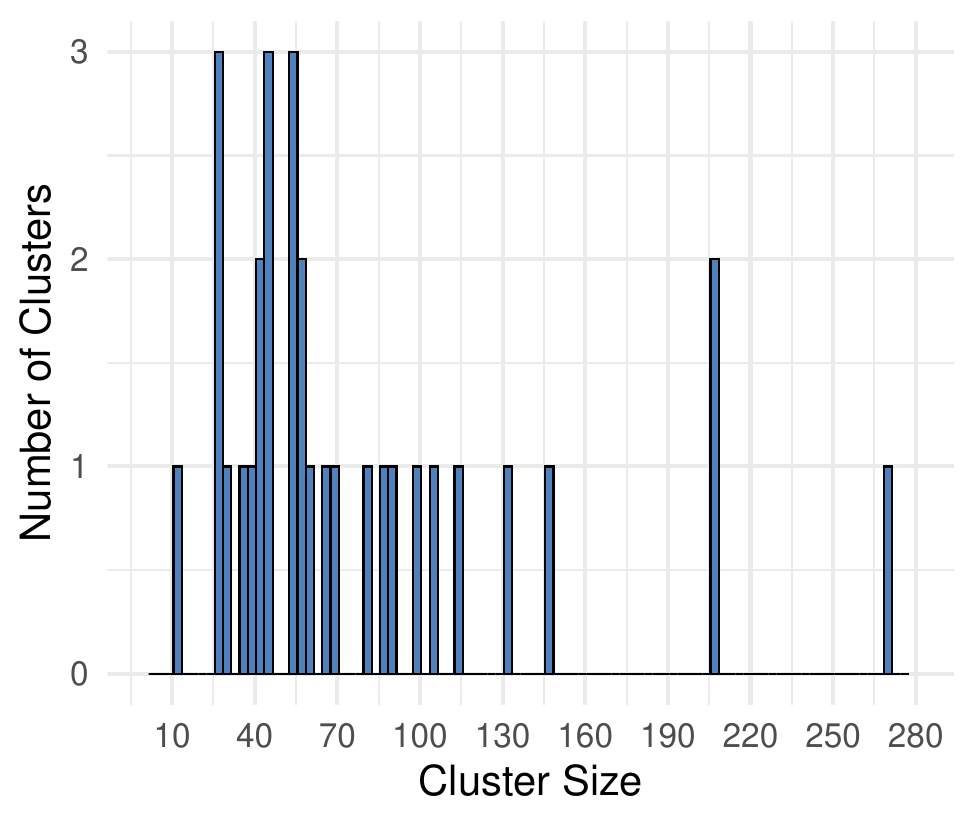}
        \caption{Distribution of Cluster Sizes}
        \label{fig:S1}
    \end{subfigure}
    \hfill
    \begin{subfigure}[t]{0.48\textwidth}
        \centering
        \includegraphics[width=\textwidth]{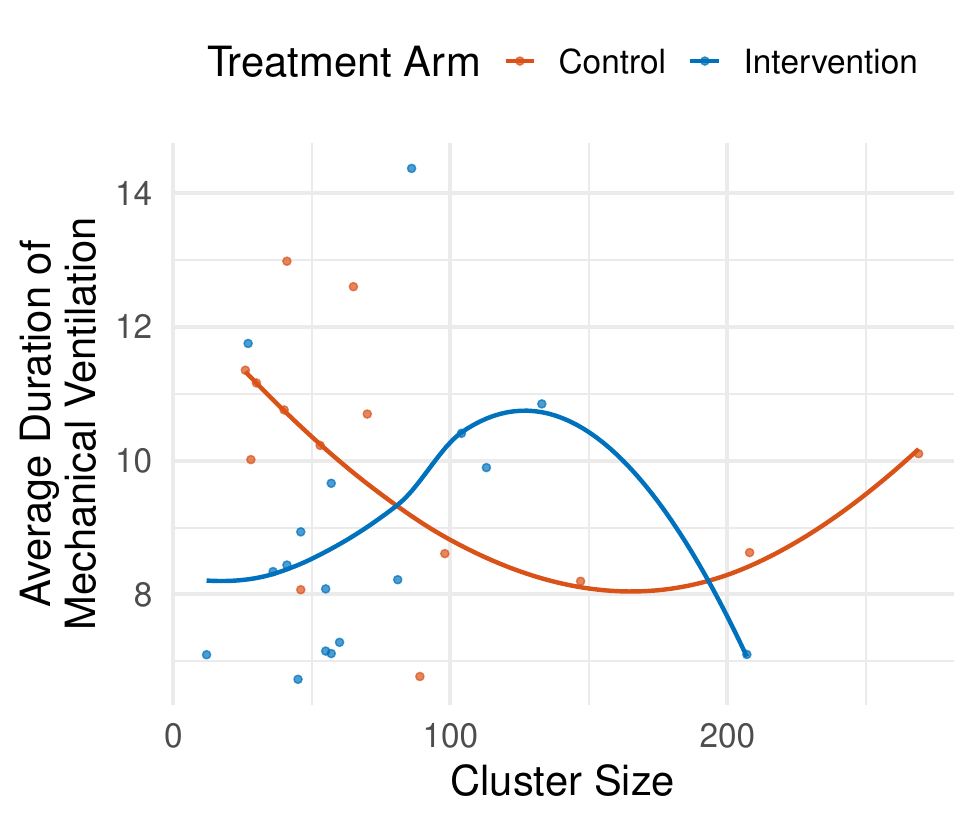}
        \caption{Average Duration of Mechanical Ventilation by Cluster Size and Treatment Arm}
        \label{fig:S2}
    \end{subfigure}
    \caption{(a) Histogram of cluster sizes; (b)   Cluster-level average duration of mechanical ventilation by cluster size and treatment arm. Each point represents a single cluster’s mean duration of mechanical ventilation, while the smooth lines represent LOESS curves fitted separately for each treatment group.}
    \label{fig:realdata}
\end{figure}

The unit of randomization in RESTORE was the pediatric intensive care unit (PICU), with 31 PICU sites across the U.S. randomized to either the intervention (sedation protocol) or usual care. 
The primary outcome of the trial was duration of mechanical
ventilation through day 28. Each PICU (cluster) enrolled between 12 and 269 patients.
Cluster sizes were highly variable, where most clusters had fewer than 100 patients but several large clusters exceeded 200 patients (Figure \ref{fig:realdata}(a)), which motivates the need to evaluate potential effect heterogeneity across cluster sizes. 
Figure \ref{fig:realdata}(b) presents the  average duration of mechanical
 ventilation for each cluster, stratified by treatment arm and plotted against cluster size. The LOESS smoothers suggest that, among smaller clusters,  the intervention arm experienced a shorter duration of mechanical ventilation compared to the control on average. However, as cluster size increases, the intervention arm tends to show longer ventilation durations relative to the control. This visible divergence in trends indicates possible modification of the treatment effect by cluster size, motivating a formal ICS hypothesis test. 

Our goal is not to reanalyze the primary treatment effect but to investigate whether treatment effect heterogeneity associated with cluster size—i.e., ICS—could compromise the validity of standard analysis methods commonly used in cluster randomized trials.
We applied the three ICS testing approaches described in Section 2 -
model-based, model-assisted, and randomization-based tests - to assess the presence of informative cluster size in this dataset. The corresponding R code is implemented in the \texttt{icstest} R package, available at https://github.com/Zhe-Chen-1999/icstest.

We first conducted tests without covariate adjustment. The model-based test examined a linear interaction between treatment assignment and cluster size, yielding a coefficient $p$-value of 0.689. The model-assisted and randomization-based tests produced $p$-values of 0.672 and 0.696, respectively. All three approaches provided no evidence of ICS in the unadjusted analysis. Next, we repeated the analyses adjusting for cluster size and the centered cluster-level risk of mortality, a covariate known to be highly predictive of the clinical outcomes. 
After covariate adjustment, the results were notably different. The model-based test yielded a $p$-value of 0.802 for the treatment-cluster size interaction, suggesting no evidence of ICS. In contrast, the model-assisted and randomization-based tests produced smaller $p$-values of 0.064 and 0.150, respectively, demonstrating potential gains in statistical power from covariate adjustment when testing for ICS. While the model-assisted test showed the strongest signal, our simulation results suggest that its Type I error may be inflated under covariate adjustment in small-sample settings, such as with 30 clusters. By comparison, the randomization-based test consistently maintained nominal Type I error across scenarios, making it a more robust and reliable choice in practice. Therefore, we did not find definitive evidence of ICS in the RESTORE study.  

\section{Discussion}
\label{s:discuss}

We seek to present a set of tests for informative cluster size derived for the CRT setting, showed that they have valid Type \RNum{1} error in a series of simulations, and described how they can be used in practice to assess statistical analysis assumptions in CRTs. The proposed model-based approach has high power, but can have inflated Type \RNum{1} error if the outcome model is misspecified. The proposed model-assisted and randomization-based tests do not require a properly specified outcome model to achieve nominal Type \RNum{1} error in the scenarios considered, but they have lower power than the model-based test in general, as a result of robustness-efficiency tradeoff. Their power can be enhanced greatly with covariate adjustment, but such power improvements will not be uniform (over an unadjusted approach) if the model is misspecified and the number of clusters is small. For CRTs with less than 60 clusters, the adjusted randomization-based test is likely the most robust approach to pursue, as it can achieve meaningful power while maintaining a nominal Type \RNum{1} error rate.

These tests can serve two immediate purposes: (i) they can be used to explore the prevalence of ICS in cluster randomized trials at large, and (ii) for trials that use mixed-effects models or GEE with exchangeable correlation structure for estimation, these tests can be used to assess the implicit assumption of no ICS which these methods make, at least in secondary analyses. However, we do not recommend that these tests are used for the purpose of prospective estimator selection. In simulations, an analyst who performs an ICS test and subsequently chooses a i-ATE estimator seems to be exercising a bias-variance tradeoff, where they will incur non-negligible bias on average under minor ICS (due to often failing to reject the null of no ICS as a Type \RNum{2} error). Notably, for an individual trial analysis, the bias-variance tradeoff is not an average of many analyses. If one fails to reject a null hypothesis of no ICS when there is truly ICS present, then the subsequent choice of estimator may result in substantial bias for a single trial.

While conditional estimands are less common targets in CRTs, there may be cases where informative cluster size corresponds to a scientific question of interest rather than a potential nuisance when targeting marginal effects, as treated in this paper. For example, one may prefer a conditional individual-average estimand that is conditional on cluster size, such that inference for an average individual from a certain cluster can be considered as a function of their cluster's size. This approach will yield its own challenges, including likely a stronger reliance on correct model specification, as well as potential ambiguity in treatment effectiveness when there is effect heterogeneity across cluster size. Furthermore, there will likely be large standard errors for CRTs with a small number of clusters. Nonetheless, investigating such heterogeneity may be important for certain interventions which have e.g., greater potential in large hospital systems but not smaller hospitals.

Overall, ICS is an underappreciated phenomenon in CRTs. Mixed-effects models and GEE with exchangeable correlation matrix are popular methods for analyzing data from CRTs which will be biased when ICS is present. Assessing the implicit assumption of no ICS that these methods make is important. Assessment of data-driven assumptions can take many forms, including graphical assessments and comparison of estimates~\citep{kahan2023informative}. The proposed hypothesis tests in this paper provide additional tools to assess this key assumption in a more formal framework.
To facilitate practical implementation of these methods, we have developed the \texttt{icstest} R package, available at https://github.com/Zhe-Chen-1999/icstest. The package implements all proposed testing procedures for informative cluster size. By making these tools readily accessible to practitioners, we hope to encourage routine assessment of this critical assumption in cluster randomized trials.

We acknowledge that other approaches to assessing ICS can be developed. For example, one could consider alternative permutation tests based on different test statistics and randomly shuffling cluster sizes to assess ICS. A formal development and comprehensive comparison to identify the globally optimal ICS test is left for future work. In addition, although our current work primarily focuses on assessing ICS on the difference scale, ICS can arise on the ratio scale for binary outcomes. For example, one may define cluster-level and individual-level risk ratio or odds ratio estimands to evaluate ICS, as discussed in \cite{kahan2024demystifying}. In such cases, a potential extension of our model-assisted test involves applying a weighted g-computation to the fitted linear model to obtain cluster-level and individual-level ratio estimates, and subsequently testing for their difference, possibly after log transformation \citep{wang2024model}. However, the design-based variance estimator for these ratio estimands would require further development and empirical validation. We view the present work as a foundational step toward testing ICS under a finite-population framework, and defer a comprehensive treatment of ratio-scale estimands and time-to-event outcomes to future research.




\section*{Acknowledgements}

Research in this article was supported by a Patient-Centered Outcomes Research Institute Award\textsuperscript{\textregistered} (PCORI\textsuperscript{\textregistered} Award ME-2022C2-27676). The statements presented are solely the responsibility of the authors and do not necessarily represent the official views of PCORI\textsuperscript{\textregistered}, its Board of Governors, or the Methodology Committee. BSB is supported by NCATS Award U24TR004437. BCK is funded by the UK Medical Research Council (grant nos. MC\_UU\_00004/07 and MC\_UU\_00004/09). \vspace*{-8pt}

\section*{Data Availability Statement}

Data access can be requested at https://biolincc.nhlbi.nih.gov/studies/restore/.


%
\bibliographystyle{unsrtnat}
\bibliography{references}

\newpage
\appendix
\section*{Web Appendix A: Tutorial for R Package icstest}
\subsection*{Package Overview and Installation}
The \texttt{icstest} package provides methods for testing \emph{informative cluster size} (ICS) in cluster randomized trials. The package implements three approaches described in the main paper: model-assisted tests, randomization-based tests, and model-based tests.


\subsubsection*{Installation}
You can install \texttt{icstest} from GitHub using \texttt{devtools}:

\begin{lstlisting}[language=R]
# Install required dependencies
install.packages(c("lmtest", "sandwich", "dplyr", "geepack"))

# Install icstest from GitHub
devtools::install_github("Zhe-Chen-1999/icstest")
\end{lstlisting}
Then load it:

\begin{lstlisting}[language=R]
library(icstest)
\end{lstlisting}

\subsection*{Core Functions}

The package provides three main testing functions:

\begin{itemize}
\item \texttt{model\_assisted\_test()}: Implements the model-assisted approach using OLS regression of
weighted cluster means
\item \texttt{randomization\_test()}: Implements the randomization-based approach using permutation testing
\item \texttt{model\_based\_test()}: Implements the model-based approach using GEE with interaction terms between treatment and cluster size
\end{itemize}

\subsection*{Example Data }
We begin with the simulated example with informative cluster size from the package documentation to demonstrate core functionality:

\begin{lstlisting}[language=R]
# Set seed for reproducibility
set.seed(123)

# Generate cluster structure
n_clusters <- 100
cluster_sizes <- sample(20:80, n_clusters, replace = TRUE)
cluster_size <- rep(cluster_sizes, cluster_sizes)
n <- sum(cluster_sizes)
cluster_id <- rep(1:n_clusters, cluster_sizes)

# Randomize treatment at cluster level
Z <- rep(rbinom(n_clusters, 1, 0.5), cluster_sizes)

# Generate outcomes with cluster size-dependent treatment effects
Y0 <- rnorm(n) + rep(rnorm(n_clusters), cluster_sizes)
tau <- rnorm(n) + rep(1.5 * (cluster_sizes > 50), cluster_sizes)
Y <- Y0 + tau * Z

# Examine the data structure
head(data.frame(Y = Y, Z = Z, cluster_id = cluster_id, 
                cluster_size = cluster_size))
\end{lstlisting}

This creates informative cluster size because larger clusters receive systematically larger treatment effects. The first few rows of data (Table~\ref{tab:cluster_data}) show the outcome, treatment, cluster ID, and cluster size for a subset of individuals in the first cluster.
\begin{table}[ht]
\centering
\caption{Sample of Simulated Cluster Data} 
\label{tab:cluster_data}
\begin{tabular}{rrrr}
  \hline
Y & Z & cluster\_id & cluster\_size \\ 
  \hline
-0.827 &    0 &    1 &   50 \\ 
  -0.928 &    0 &    1 &   50 \\ 
  -1.532 &    0 &    1 &   50 \\ 
  -0.626 &    0 &    1 &   50 \\ 
  -1.366 &    0 &    1 &   50 \\ 
  -2.249 &    0 &    1 &   50 \\ 
   \hline
\end{tabular}
\end{table}

\subsection*{Running the Tests}

\subsubsection*{Model-Assisted Test}
The model-assisted test uses OLS regression of
weighted cluster means, where weights are proportional to deviations of cluster size from the average. This test uses the function \texttt{model\_assisted\_test()}, which takes (minimally) the outcome \(Y\), the treatment indicator \(Z\), and the cluster IDs.

\begin{verbatim}
model_assisted_test(Y, Z, cluster_id, covariates = NULL, adjust_size = TRUE)
\end{verbatim}

\textbf{Arguments:}
\begin{itemize}
\item \texttt{Y}: Numeric vector of individual-level outcomes. Must have same length as \texttt{Z} and \texttt{cluster\_id}.
\item \texttt{Z}: Numeric vector of individual-level treatment assignments (0/1). Binary treatment indicator.
\item \texttt{cluster\_id}: Vector identifying cluster membership. Each unique value represents one cluster.
\item \texttt{covariates}: Optional matrix or data.frame of cluster-level covariates. Number of rows must equal number of clusters. Columns are automatically mean-centered.
\item \texttt{adjust\_size}: Logical (default \texttt{TRUE}). When \texttt{TRUE}, includes cluster size as a covariate and its interaction with treatment.
\end{itemize}

\textbf{Returns:} List with method name, test statistic and $p$-value.

\textbf{Example:}
\begin{lstlisting}[language=R]
# Run model-assisted test
ma_result <- model_assisted_test(Y, Z, cluster_id)

# Display results
print(ma_result)

# Expected output:
# $method
# [1] "Model-Assisted Test"
# 
# $t_stat
# [1] 2.419371
# 
# $p_value
# [1] 0.0174
\end{lstlisting}
The $p$-value of 0.0174 provides strong evidence against the null hypothesis of no informative cluster size.

\subsection*{Randomization-Based Test}

The randomization-based test permutes the cluster-level treatment assignments and computes the proportion of permuted test statistics that are as or more extreme than the observed. The function \texttt{randomization\_test()} performs a randomization-based test based on \texttt{model\_assisted\_test()}.

\begin{verbatim}
randomization_test(Y, Z, cluster_id, covariates = NULL, 
                   adjust_size = TRUE, n_perms = 5000)
\end{verbatim}

\textbf{Arguments:}
\begin{itemize}
\item \texttt{Y}, \texttt{Z}, \texttt{cluster\_id}, \texttt{covariates}, \texttt{adjust\_size}: Same as \texttt{model\_assisted\_test}
\item \texttt{n\_perms}: Integer (default 5000). Number of permutations for randomization test. Higher values provide more precise p-values but increase computation time.
\end{itemize}


\textbf{Return Value:} List with components:
\begin{itemize}
\item \texttt{method}: Character string ``Randomization-Based Test"
\item \texttt{statistic}: Observed test statistic
\item \texttt{p\_value}: Permutation-based $p$-value
\end{itemize}

\textbf{Example:}
\begin{lstlisting}[language=R]
# Run randomization-based test with 5000 permutations
rand_result <- randomization_test(Y, Z, cluster_id, n_perms = 5000)

# Display results
print(rand_result)

# Expected output:
# $method
# [1] "Randomization-Based Test"
# 
# $statistic
# [1] 2.419371
# 
# $p_value
# [1] 0.019
\end{lstlisting}

The resulting $p$-value is close to that of the model-assisted test, suggesting we can reject the null
hypothesis at conventional significance levels. One may choose the number of permutations (\texttt{n\_perms}) to balance precision vs computation time.

\subsection*{Model-Based Test}

If you want to explicitly test for an interaction of cluster size with treatment (i.e. check whether treatment effect changes with a specific functional form of cluster size), you can use \texttt{model\_based\_test()}. It uses a formula interface (as in R’s modeling functions) and data frame input. 

\begin{verbatim}
model_based_test(model_formula, interaction_term_name, model_data, cluster_id)
\end{verbatim}

\textbf{Arguments:}
\begin{itemize}
\item \texttt{model\_formula}: R formula object specifying the GEE model, including interaction between treatment and cluster size.
\item \texttt{interaction\_term\_name}: Character string specifying exact name of the treatment-cluster size interaction term (e.g., \texttt{"Z:cluster\_size"}).
\item \texttt{model\_data}: Data frame containing all variables in \texttt{model\_formula}: outcome, treatment, cluster size, cluster ID, and any covariates to be adjusted for.
\item \texttt{cluster\_id}: A vector or column in \texttt{model\_data} indicating the cluster membership of each unit.
\end{itemize}

\textbf{Returns:} Numeric $p$-value for the specified interaction term.

\textbf{Example:}
\begin{lstlisting}[language=R]
# Prepare data frame for model-based test
data <- data.frame(
  Y = Y,
  Z = Z,
  cluster_size = cluster_size,
  cluster_id = cluster_id
)

# Test linear interaction between treatment and cluster size
formula_linear <- Y ~ Z * cluster_size
mb_pval_linear <- model_based_test(
    model_formula = formula_linear, 
    interaction_term_name = "Z:cluster_size", 
    model_data = data, 
    cluster_id = cluster_id
)

# Test logarithmic interaction between treatment and cluster size
formula_log <- Y ~ Z * log(cluster_size)
mb_pval_log <- model_based_test(formula_log, "Z:log(cluster_size)",
                                data, cluster_id) 

# Test threshold interaction (our actual simulation structure)
data$large_cls <- as.numeric(data$cluster_size > 50)
formula_threshold <- Y ~ Z * large_cls
mb_pval_threshold <- model_based_test(formula_threshold, "Z:large_cls",
                                data, cluster_id) 
                                
# Compare different functional forms
interaction_comparison <- data.frame(
  Interaction_Type = c("Linear", "Logarithmic", "Threshold"),
  P_value = c(mb_pval_linear, mb_pval_log, mb_pval_threshold),
  Formula = c("Z * cluster_size", "Z * log(cluster_size)", "Z:large_cls")
)
print(interaction_comparison)

# Interaction_Type      P_value               Formula
# 1           Linear 7.139569e-04      Z * cluster_size
# 2      Logarithmic 1.554608e-03 Z * log(cluster_size)
# 3        Threshold 8.683351e-06           Z:large_cls
\end{lstlisting}

Here \texttt{Z * cluster\_size}, \texttt{Z * log(cluster\_size)}, and \texttt{Z:large\_cls} indicate which coefficient in the model corresponds to the treatment × cluster-size interaction to test. All specifications detect strong evidence for ICS. 
In our simulation, the threshold model should perform best since we explicitly programmed a step function at cluster size 50. In real data, analysts are encouraged to explore multiple functional forms of cluster size for robustness.

\subsection*{Covariate Adjustment}
To better reflect applied settings, we generate baseline cluster-level covariates that may be prognostic of the outcome. These will later be used to illustrate how covariate adjustment is handled by each test. 

\begin{lstlisting}[language=R]
# Generate cluster-level covariates
mortality_risk <- rnorm(n_clusters, 0.2, 0.1)
hospital_size <- sample(c("Small", "Medium", "Large"), n_clusters, 
                       replace = TRUE, prob = c(0.4, 0.4, 0.2))
hospital_size_num <- as.numeric(factor(hospital_size, 
                                levels = c("Small", "Medium", "Large")))

# Create covariate matrix
covariates <- data.frame(
  mortality_risk = mortality_risk,
  hospital_size = hospital_size_num
)
\end{lstlisting}
In practice, \texttt{mortality\_risk} could represent baseline severity of patients in a hospital, while \texttt{hospital\_size} categorizes institutions by capacity. Incorporating such covariates into ICS tests may increase power and provide more efficient inference. 
All three testing procedures in the \texttt{icstest} package can adjust for baseline covariates:
\begin{lstlisting}[language=R]
# Model-assisted test with covariate adjustment
ma_adj <- model_assisted_test(Y, Z, cluster_id, 
                                      covariates = covariates, 
                                      adjust_size = TRUE)
print(ma_adj)

# Expected Output:
# $method
# [1] "Model-Assisted Test"
# 
# $t_stat
# [1] 3.077903
# 
# $p_value
# [1] 0.002746957

# Randomization test with covariate adjustment
rand_adj <- randomization_test(Y, Z, cluster_id, 
                                     covariates = covariates,
                                     adjust_size = TRUE,
                                     n_perms = 5000)
print(rand_adj)

# Expected Output:
# $method
# [1] "Randomization-Based Test"
# 
# $statistic
# [1] 3.077903
# 
# $p_value
# [1] 0.0032

# Model-based test with covariate adjustment
# Add covariates to model data
model_data_adj <- cbind(data, 
                    mortality_risk = rep(mortality_risk, cluster_sizes),
                    hospital_size = rep(hospital_size_num, cluster_sizes))

# Multiple interaction terms
formula_adj <- Y ~ Z + cluster_size + mortality_risk + hospital_size + 
  Z:cluster_size + Z:mortality_risk + Z:hospital_size

mb_adj <- model_based_test(formula_adj, "Z:cluster_size",
                                     model_data_adj, cluster_id)

cat("Adjusted model-based test p-value:", mb_adj, "\n")
# Adjusted model-based test p-value: 4.268617e-05 
\end{lstlisting}

The adjusted results reveal smaller $p$-values and strengthen evidence for ICS across all three 
testing approaches.

\subsection*{Reference to Main Paper}

For full details on the operating characteristics of these tests (Type I error, power) under varying magnitudes of ICS and numbers of clusters, see Section 3 of the main manuscript.

\subsection*{Package Citation}

When using the \texttt{icstest} package, cite both the package and the methodology paper:

\begin{verbatim}
# R citation
citation("icstest")

# Paper citation
# Blette, B.S., Chen, Z., Kahan, B.C., Forbes, A., Harhay, M.O., 
# and Li, F. (2025). "Evaluating informative cluster size in 
# cluster randomized trials." 
\end{verbatim}

\section*{Web Appendix B: Supplemental Simulations under ICC = 0.01}

In this section, we present results from simulations that mirror Scenarios 1-3, generating data such that the true ICC is equal to 0.01 rather than 0.1 as reported in the main manuscript. All other data-generating procedures remain the same. Figures 5-7 present the results for Scenario 1, 2, and 3, respectively. All results are similar to those presented in the main manuscript. Thus, the tests continue to be valid under ICC = 0.01. In Scenarios 1 and 2, the supplemental simulations yielded slightly higher power for the proposed tests, as expected with a lower ICC.

\begin{sidewaysfigure}
    \centering
    \includegraphics[width=1\linewidth]{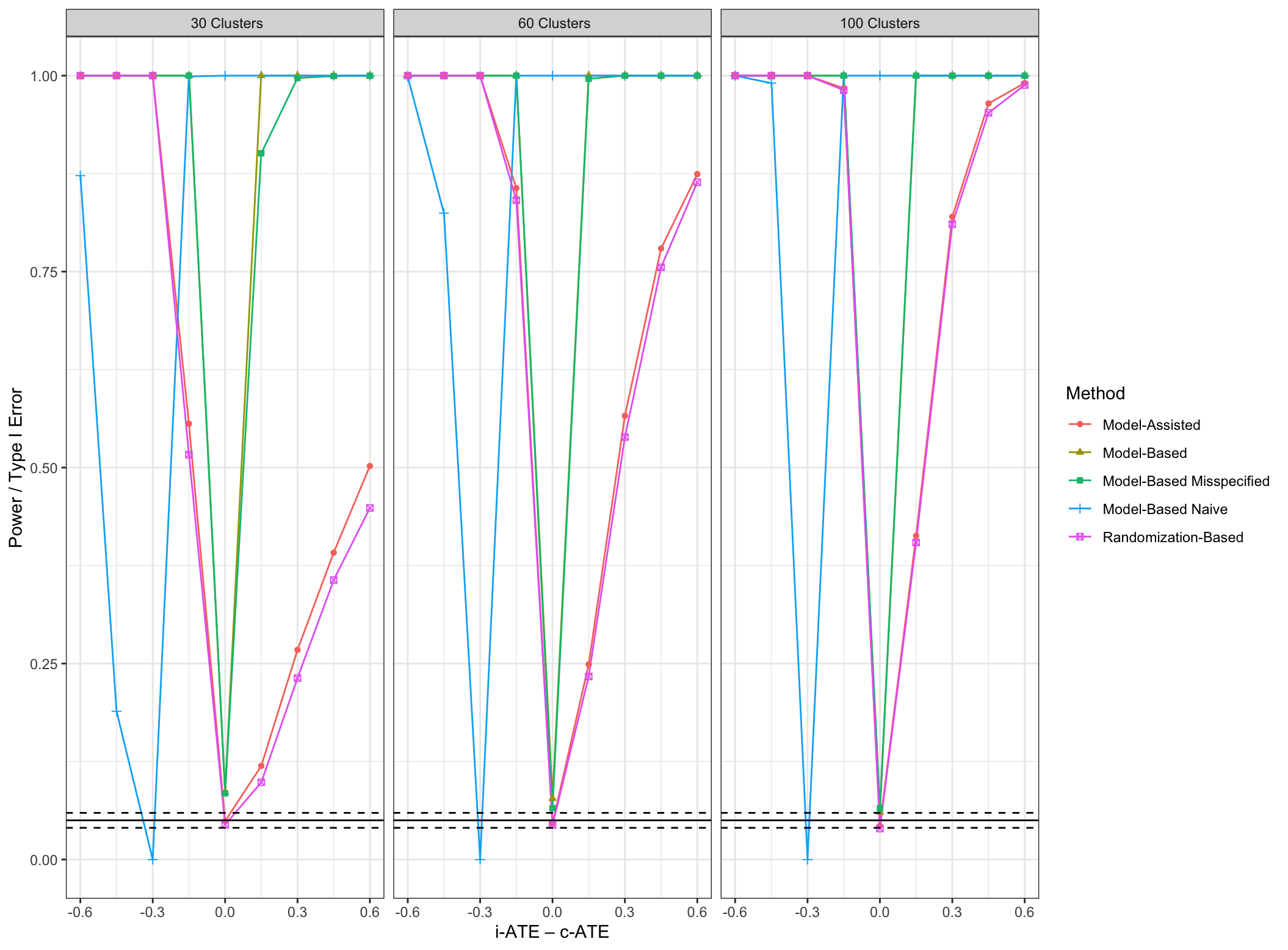}
    \caption{Results from Simulation 1 as described in the main manuscript but with ICC = 0.01. The dashed lines indicate Monte Carlo standard error for 2000 simulations assuming a true Type I error of 5\%; this range is relevant for the middle of each panel where the null hypothesis holds.}
    \label{fig:fig1a}
\end{sidewaysfigure}

\begin{sidewaysfigure}
    \centering
    \includegraphics[width=1\linewidth]{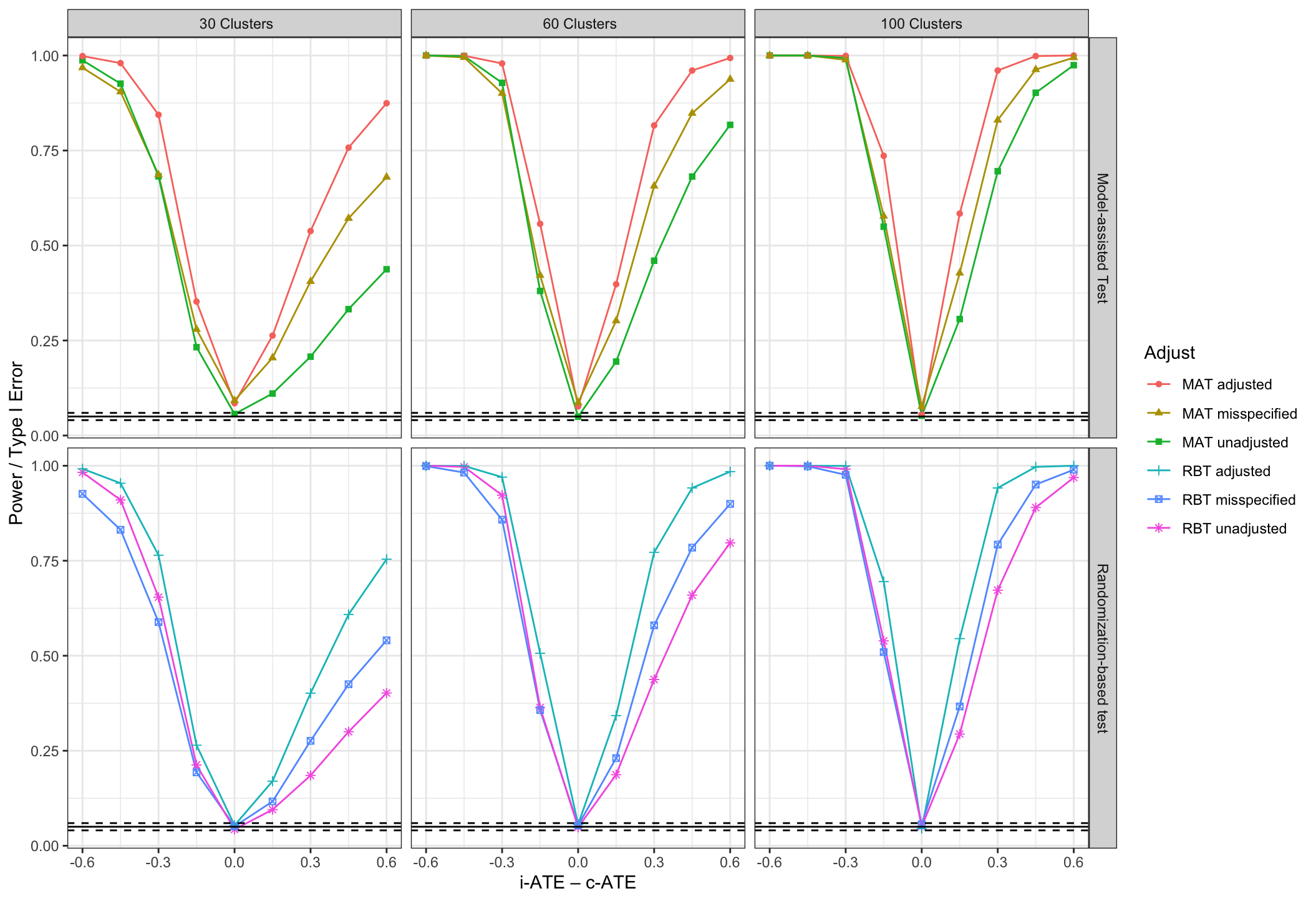}
    \caption{Results from Simulation 2 with ICC = 0.01. MAT = model-assisted test. RBT = randomization-based test. The dashed lines indicate Monte Carlo standard error for 2000 simulations assuming a true Type I error of 5\%; this range is relevant for the middle of each panel where the null hypothesis holds.}
    \label{fig:fig2a}
\end{sidewaysfigure}

\begin{sidewaysfigure}
    \centering
    \includegraphics[width=1\linewidth]{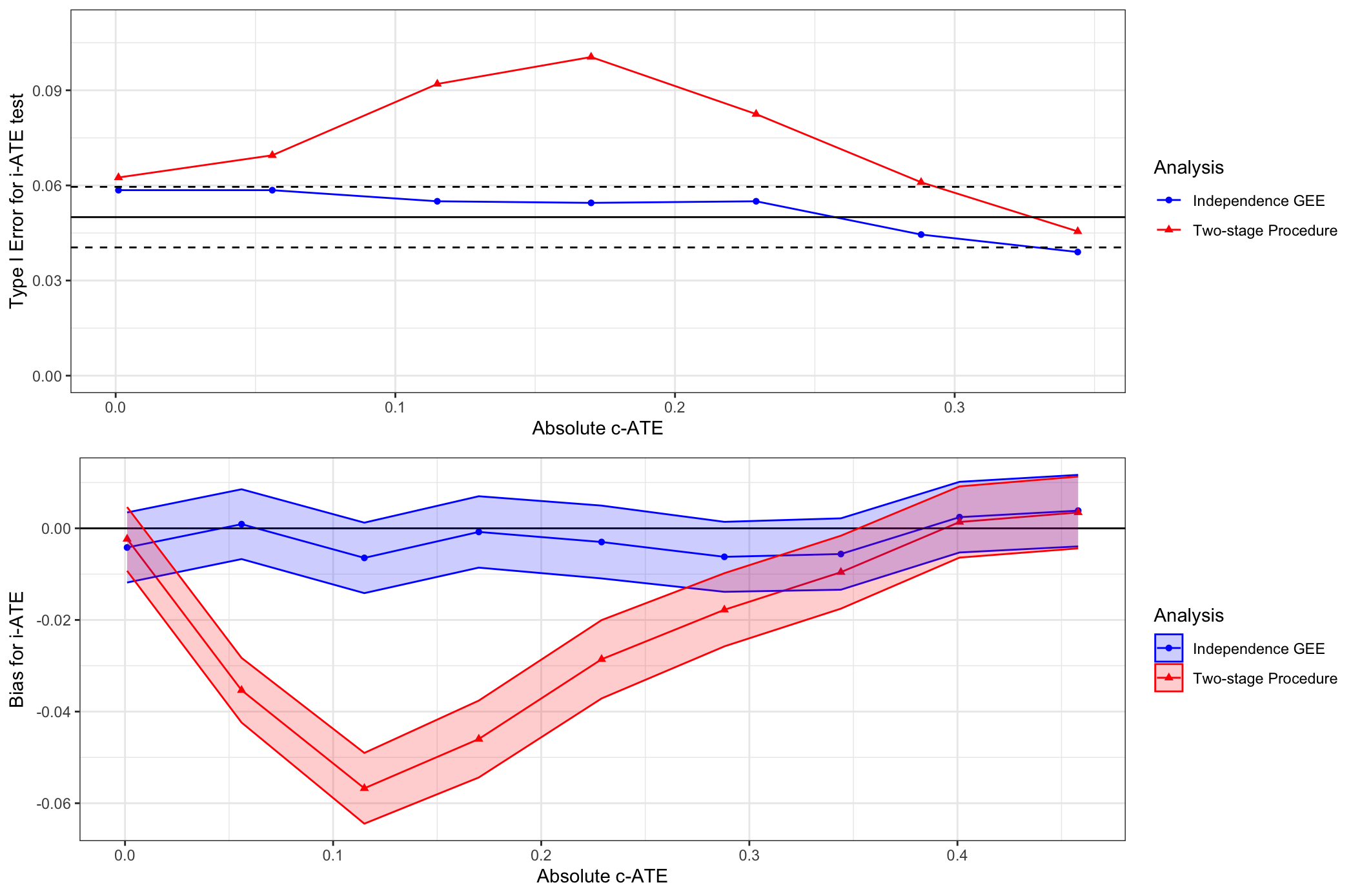}
    \caption{Results from Simulation 3 comparing an analyst who performs a two-stage procedure to an analyst who uses independence estimating equations, now under ICC = 0.01. In the top panel, Type I error is reported and the dashed lines indicate Monte Carlo standard error for 2000 simulations assuming a true Type I error of 5\%. In the bottom panel, bias is reported with corresponding Monte Carlo confidence intervals to quantify uncertainty from using a finite number of simulations.}
    \label{fig:fig3a}
\end{sidewaysfigure}

\section*{Web Appendix C: Supplemental Simulations to Investigate Type I Error}

In this section, we present results from an extension of Scenario 2. We only consider the data generating mechanisms which result in no ICS (such that the null hypothesis holds), but we vary the number of clusters more granularly to develop a deeper understanding of when nominal Type I error can be expected. In particular, these are scenarios with $k=5$ as described in the main manuscript, but where the number of clusters is varied from 30 to 100 in increments of 10. An ICC = 0.1 was specified.

Results are presented in Figure 8. Dashed lines indicate the Monte Carlo simulation error bounds for a Type I error parameter equal to 0.05. We ran 1,000 simulation iterations. In the scenarios considered, the adjusted model-assisted test required around 70 clusters to achieve an acceptable Type I error rate. Thus, it may not be suitable for smaller CRTs. In contrast, the unadjusted model-assisted test and all randomization-based tests achieved appropriate Type I error control across the full range of number of clusters considered. In conjunction with the power results from Scenario 2, the adjusted randomization-based test is likely a good choice in practice for CRTs with less than 70 clusters.

\begin{sidewaysfigure}
    \centering
    \includegraphics[width=0.8\linewidth]{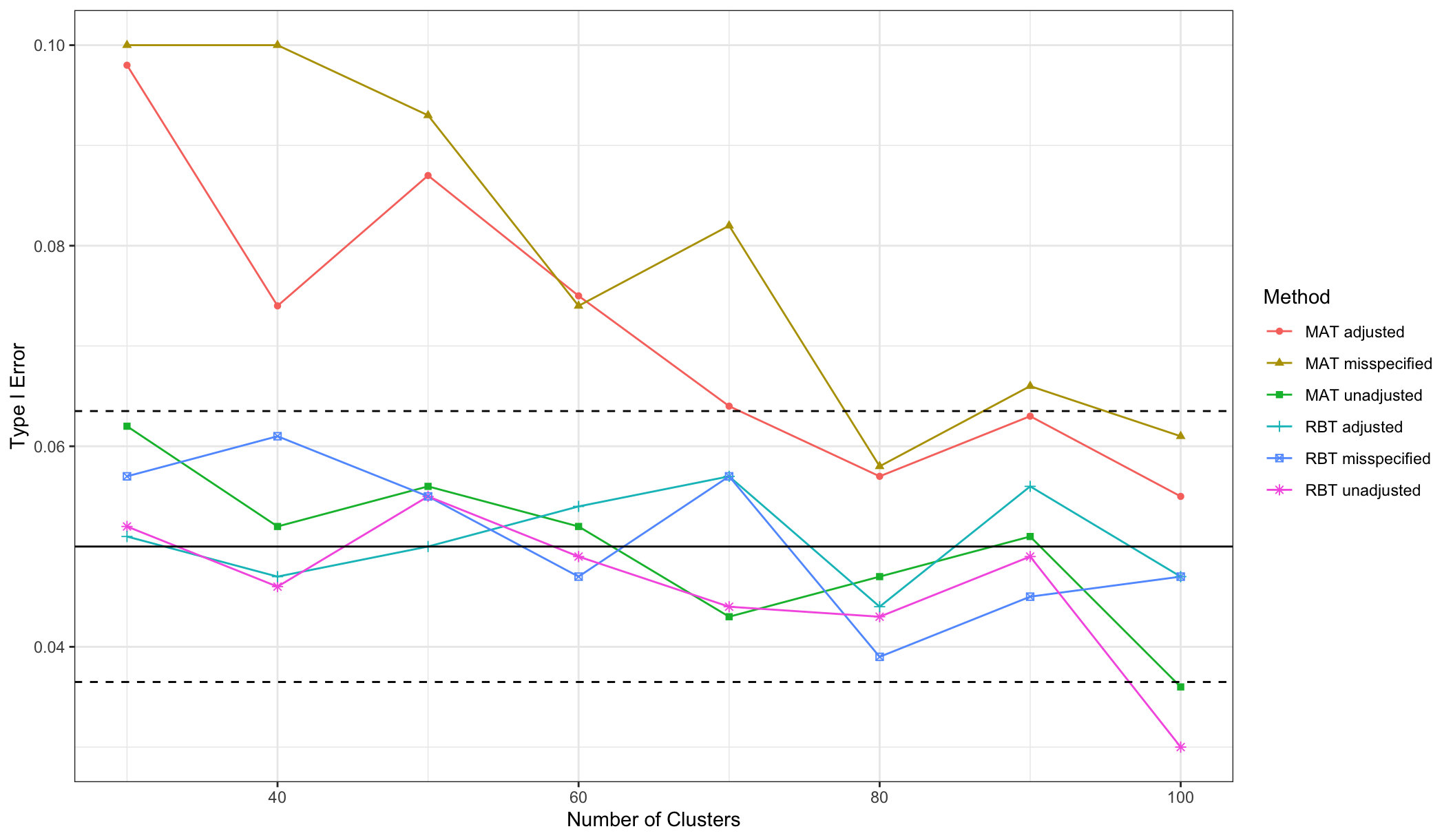}
    \caption{Results from an extension of Scenario 2 to investigate Type I error across number of clusters. MAT = model-assisted test. RBT = randomization-based test. The dashed lines indicate Monte Carlo standard error for 1000 simulations assuming a true Type I error of 5\%.}
    \label{fig:fig4a}
\end{sidewaysfigure}

\end{document}